\documentclass[preprint,superscriptaddress,nofootinbib, notitlepage, double-spaced, showpacs,floatfix,longbibliography]{revtex4}
\usepackage{epsfig}
\usepackage{amsthm}
\usepackage{amsfonts}
\usepackage{float}
\usepackage{amsmath,amssymb}
\usepackage{color}
\usepackage[colorlinks = true,
linkcolor = blue,
urlcolor  = blue,
citecolor = red,
anchorcolor = green]{hyperref}
\usepackage{graphicx}
\usepackage{dcolumn}
\usepackage{bm}
\usepackage{hyperref}

\newcommand{\nn}{\nonumber}
\newcommand{\be}{\begin{equation}}
\newcommand{\ee}{\end{equation}}
\newcommand{\bea}{\begin{eqnarray}}
\newcommand{\eea}{\end{eqnarray}}

\begin{document}
\title {Nonlinear waves in a hot, viscous and non-extensive quark-gluon plasma}
\author{Golam Sarwar}
\email{golamsarwar1990@gmail.com}
\affiliation{School of Physical Sciences, National Institute of Science Education and Research, HBNI, Jatni-752050, Odisha, India} 
\affiliation{Department of Physics,University of Calcutta, 92, A.P.C. Road, Kolkata-700009, India}
\author{Md Hasanujjaman}
\email{jaman.mdh@gmail.com}
\affiliation{Department of Physics, Darjeeling Government College, Darjeeling- 734101, India}
\author{Trambak Bhattacharyya}
\email{trambak.bhattacharyya@gmail.com }
\affiliation{Bogoliubov Laboratory of Theoretical Physics, Joint Institute for Nuclear Research, Dubna 141980, Moscow Region, Russian Federation} 
\author{Mahfuzur Rahaman}
\email{mahfuzurrahaman01@gmail.com }
\affiliation{Department of Physics, Darjeeling Government College, Darjeeling- 734101, India}
\affiliation{Variable Energy Cyclotron Centre, 1/AF Bidhan Nagar, Kolkata- 700064, India}
\affiliation{Homi Bhabha National Institute, Training School Complex, Mumbai - 400085, India}

\author{Abhijit Bhattacharyya}
\email{abhattacharyyacu@gmail.com}
\affiliation{Department of Physics,University of Calcutta, 92, A.P.C. Road, Kolkata-700009, India}
\author{Jan-e Alam}
\email{jane@vecc.gov.in}
\affiliation{Variable Energy Cyclotron Centre, 1/AF Bidhan Nagar, Kolkata- 700064, India}
\affiliation{Homi Bhabha National Institute, Training School Complex, Mumbai - 400085, India}

\def\zbf#1{{\bf {#1}}}
\def\bfm#1{\mbox{\boldmath $#1$}}
\def\hf{\frac{1}{2}}
\def\sl{\hspace{-0.15cm}/}
\def\omit#1{_{\!\rlap{$\scriptscriptstyle \backslash$}
{\scriptscriptstyle #1}}}
\def\vec#1{\mathchoice
        {\mbox{\boldmath $#1$}}
        {\mbox{\boldmath $#1$}}
        {\mbox{\boldmath $\scriptstyle #1$}}
        {\mbox{\boldmath $\scriptscriptstyle #1$}}
}
\def \beq{\begin{equation}}
\def \eeq{\end{equation}}
\def \beqa{\begin{eqnarray}}
\def \eeqa{\end{eqnarray}}
\def \pd{\partial}
\def \nn{\nonumber}
\begin{abstract}
The effects of the non-extensive statistics on 
the nonlinear propagation of perturbations have been studied within the scope of relativistic second 
order dissipative hydrodynamics with the non-extensive equation of state. We have shown that the equations,  
describing the  propagation of nonlinear waves under such situation are KdV-type (Korteweg-De Vries). 
Apart from their preserved solitonic behaviour the dissipative nature of these waves are also observed. 
The waves with larger amplitude and width dissipate less and propagate faster and these waves deplete more for
both smaller values of Tsallis parameter ($q$) and temperature ($T$) of the medium. For vanishingly small transport 
coefficients the nonlinear waves show breaking nature. These findings suggest that the nature of the 
propagation of the nonlinear waves may serve as a good probe to differentiate between the extensive and 
non-extensive  thermodynamic nature  of a fluid, such as the quark-gluon plasma, produced in relativistic 
nuclear collisions.  
\end{abstract}

\maketitle
\section{Introduction}
\label{sec1}
The propagation of perturbations in fluids have been widely used to unveil  the thermodynamic state of 
the fluid.  Depending on the magnitude of the  perturbation, 
it has been categorized as linear~\cite{pub.1002429324,Lick1990} or nonlinear. 
If the magnitude of the perturbation is small compared to the average
value of the corresponding physical quantities then it is categorized as linear.
On the other hand, if the perturbation is comparable to the corresponding
average value then it has to be treated as nonlinear.
In the context of  quark-gluon plasma
(QGP), formed in relativistic heavy-ion collisions (RHICs) and Large Hadron Collider (LHC), 
the response of   QGP to the perturbations may help in
characterizing the medium. The locally equilibrated system of QGP 
responds to its high internal pressure by rapid expansion against the vacuum. 
The  QGP formed in the violent collisions of nuclei 
cools due to expansion and subsequently transforms to hadronic matter  
via a cross over transition or through an intermediary coexisting phase 
(of QGP and hadrons) depending on
the value of temperature ($T$) and baryonic chemical potential ($\mu$) attained 
in these collisions.
When these hadrons cease to interact, their momenta
get frozen to certain value and finally move freely to
the detector.  The momentum spectra of these detected hadrons are analyzed
to understand the dynamics of the collisions.
In addition to the hadrons, penetrating probes, such as photons and 
lepton pairs emanated
from each space time point of the medium, are also detected to
shed light on the early hot and dense phase of QGP~\cite{Busza:2018rrf}.

The pressure gradient in QGP can be anisotropic and the degree of anisotropy depends 
on the value of  impact parameter of the collisions. 
An anisotropic interacting system of quarks and gluons created after the
collisions will develop a pressure gradient of different 
magnitudes along different directions. This anisotropic 
pressure gradient will result in different expansion rate which
in turn will induce anisotropy in momentum distribution of the
hadrons originating from the system on hadronization. For the small
initial spatial anisotropy, the produced momentum anisotropy  could be interpreted as linear response to the initial
spatial  anisotropy. However, nonlinear effects will be crucial 
if the  anisotropy in  initial geometry is large
due to collisions at large impact parameter.
Moreover, quarks and gluons produced in the early
stage of collision with high momenta, 
do not equilibrate but propagate 
as jets through the thermal medium created due to the rescattering of the 
low momentum quarks and gluons. These 
jets, while propagating, deposit energy in the medium~\cite{Shuryak:2009cy,Betz:2009jw}. 
If the magnitude of the energy density resulting from the jet-medium 
interaction is  comparable to the energy density of the
medium, then the effects of jet propagation on the medium needs to  be treated as nonlinear. 
On the other hand, smaller value of the deposited energy density  allows
the study of the jet-medium interaction within the domain of linear response theory. 
The propagation of linear perturbations in QGP has been studied extensively 
by several authors~\cite{ Staig:2010pn,Staig:2011wj,Sarwar:2015mma,Rafiei:2016zxk,Hasanujjaman:2020zmn,Hasanujjaman:2020zex,Minami:2009hn,Rahaman:2017ezf}. 
However, the nonlinear aspects of the perturbations in QGP have been studied only
by a few authors~\cite{Raha:1983pr,Fowler:1982ygc,Raha:1982ex,Hefter:1984pu, Fogaca:2009wf,Fogaca:2014gwa,Sarwar:2020oux}. 
In Refs.~\cite{Raha:1983pr,Fowler:1982ygc} the authors have considered the background medium to be an  
 ideal fluid and found that the nature of perturbations is solitonic. 
Fogaca \emph {et. al.} in Refs.~\cite{Fogaca:2009wf,Fogaca:2014gwa} went further and introduced 
the shear viscous effects in the medium and considered the perturbations up to 
second order. They too, found the solitonic nature of the perturbations.
In these studies, the local equilibrium is treated
within the scope of Gibbs-Boltzmann (GB) extensive distribution.
However, particle spectra from small systems produced in $p+p$ collision are 
found to be well described by 
non-extensive equilibrium distribution suggested by 
Tsallis~\cite{Tsallis:1987eu,Tsallisbook,CMS:2010tjh,ALICE:2017xrp,Biro:2008hz,Cleymans:2013rfq,Marques:2015mwa,
Bhattacharyya:2015hya,Tripathy:2016hlg,Grigoryan:2017gcg,Bhattacharyya:2017hdc,Azmi:2019irb}. 
In fact, the  data is better reproduced~\cite{Wong:2015mba,Azmi:2015xqa,Bhattacharyya:2017hdc, Cleymans:2012ya,Zheng:2015mhz} 
with Tsallis distribution than the GB distribution. 

The appearance of the  entropic index, $q\neq 1$ (also called non-extensive parameter) in Tsallis 
distribution is attributed to the local fluctuations in the system~\cite{Wilk:1999dr}. It is important
to recall at this point that, in the limit of $q\rightarrow 1$, the Tsallis distribution
approaches the GB distribution.
In principle, $q$ can be determined from the dynamics of the system. For some cases,  
like non-linear Fokker-Planck equation \cite{Tsallis_1996}, Boltzmann lattice model \cite{PhysRevE.68.025103}, 
cold atoms in optical lattices \cite{Douglas:2006zz} etc,  the 
$q$ is known analytically in terms of microscopic or mesoscopic quantities. 
A relation between the entropic index $q$ and the
basic parameters of the QCD has been established
in Refs.~\cite{Deppman:2020gbu,Deppman:2019yno}. Furthermore, it was shown that, this value
of $q$ can be used to  reproduce the transverse momentum spectra of hadrons 
produced in relativistic proton+proton collisions~\cite{Deppman:2020jzl}. 
The value of $q$ for a many body system
has been expressed in terms of the number of particles ($N$) as,
$q=1+2N^{-1}/3$ and it was shown that, the two particle correlation becomes 
weaker 
as $N\rightarrow\infty$ {\it i.e.} when the system
approaches  the domain of molecular chaos~\cite{2020PhRvE.101d0102L}. 
This nicely depicts the importance of non-extensitivity
of a system with finite number of particles and 
has relevance for system formed in high energy collisions of protons and nuclei.  
It has been shown in Ref.~\cite{Wilk:1999dr} that, in a system with fluctuating temperature 
zones, 
$q$ is related to the relative variance in temperature. 
The interplay between the parameter $q$  and
the QCD dynamics  has been studied in Refs.~\cite{Rozynek:2016ykp,Rozynek:2018tev}.
In relativistic nuclear collisions, the $q$ 
is treated as a parameter to fit the  
transverse momentum spectra of hadrons,
which can in principle vary from 0 to $\infty$. However,
for the phenomenological studies, the values of $q$  are generally greater than
one and from the thermodynamic considerations it has an upper bound of 4/3 \cite{Bhattacharyya:2016lrk}.
The descriptions of different experimental observables, which 
carry the effect of fluctuations, require $q\neq 1$ as argued 
in Refs.~\cite{Wilk:1999dr,Navarra:2003am,Navarra:2003um,Biyajima:2004ub,Biyajima:2006mv,Wilk:2000yv,Wilk:2009nn,Biro:2004qg}. 
The value of  $q$ varies with collision centrality, energy and system size~\cite{Biyajima:2004ub,Biyajima:2006mv}. 
It has been reported that the system size alone is not sufficient to explain the value of 
non-extensive parameter~\cite{Deb:2019yjo}. 

Moreover, there are other observables advocated for choosing the 
non-extensive local thermal equilibrium. The system size dependence 
of quantum fluctuations has been studied theoretically in 
Refs.~\cite{Das:2020ddr,Das:2021aar,Das:2021rck}. These fluctuations 
may be represented by the $q$ parameter. 
The particle spectra for small systems depends on the 
$q-$parameter. A similar situation may arise in  
bigger systems also as the spatial inhomogeneity 
created in such systems (produced in nuclear collisions)
may contain some smaller zones  
of size similar to that of the entire system produced in a $p+p$ 
collision. This indicates that the local equilibrium 
of such systems may be considered to be non-extensive in nature.  At this stage one should 
find out ways to  distinguish 
whether non-extensive local equilibrium is the underlying nature 
 or an extensive local equilibrium is the correct 
situation for the fluid. This requires investigations of 
the effect of non-extensivity on the response of the fluid 
to the perturbations. 
 
To understand the distinguishing effects between the non-extensivity
end extensivity in locally equilibrated fluid, one considers 
two aspects in this regard:
a) the equation of state to be of a non-extensive form, 
and b) the fluid dynamic equations should also be of  
non-extensive type, {\it{i.e.}}, to say that the fluid velocity 
must also be defined by taking into account 
the non-extensivity, which will clearly be different from the 
extensive one~\cite{Osada:2008sw}. 

Recently, the study of the propagation of non-linear waves has 
considered only the first order deviation from the extensive one, 
where the non-linear wave equation for first-order perturbation 
has been derived for the ideal fluid background with fixed thermodynamic 
quantities~\cite{Bhattacharyya:2020sua}. 
In Ref.~\cite{Bhattacharyya:2020sua} 
breaking wave solution has been obtained for non-dissipative propagation of the  
nonlinear waves. 
However, it has been shown in 
Ref.~\cite{Osada:2008sw} that the ideal non-extensive 
hydrodynamics ($q$-hydrodynamics) is equivalent to extensive 
dissipative hydrodynamics ($d$-hydrodynamics). The corresponding 
dissipation should result from the non-extensive nature of fluid 
velocity field~\cite{Osada:2008sw}. Therefore, even in ideal 
non-extensive background, one expects dissipative propagations 
of perturbations. Moreover, the local equilibrium leads to inherent 
gradients in the dynamical fields of fluid which leads to transport 
phenomenon. Therefore, it is important to investigate the 
dissipative propagation of   nonlinear waves in the 
non-extensive background to understand the role of non-extensivity 
on the propagation of perturbations. 

In this work, we proceed to investigate the propagation of nonlinear 
waves in QGP with non-extensive equation of state (EoS) and 
also with viscous-coefficients  appearing from non-extensive nature of fluid. The aim is to find how 
the conclusion changes with the change in background (either extensive or 
non-extensive) and how $q$ affects the propagation of the 
nonlinear waves. The required equations for propagation (of 
nonlinear waves) are derived from M\"{u}ller-Israel-Stewart (MIS) hydrodynamics and the order of perturbations are considered up to the second order.

The paper is organized as follows:  in the next section, {\it{i.e} } in Sec.~\ref{sec2}, we derive the required equations for nonlinear 
waves from 
MIS theory with all the relevant transport coefficients.  
In the subsection \ref{sec2}.A we discuss the MIS 
relativistic causal hydrodynamics. Subsection \ref{sec2}.B contains 
the equation of motion of the fluid in (1+1) dimension. 
In subsection \ref{sec2}.C, the non-linear equations governing the propagation of the perturbations
through the fluid have been provided.  
The Equation of State (EoS) for the non-extensive 
fluid has been discussed in Sec.~\ref{sec3}.
We present our results in Sec.~\ref{sec4} and finally summarize the findings 
in Sec.~\ref{sec5}. Some of the mathematical equations and expressions are presented in 
Appendix \ref{appendixA}, \ref{appendixB} and  \ref{appendixC}.
We have used
natural units ($\hbar=c=1, k_B=1$) and the signature of the Minkowski metric 
as:  $g^{\mu \nu}=(1,-1,-1,-1)$.


\section{Framework of relativistic viscous hydrodynamics}
\label{sec2}
In the following subsections we derive the equation of motions 
of non-linear perturbations propagating through a relativistic viscous
fluid. 

\subsection{Hydrodynamic equations}

The non-extensive version of ideal and first order dissipative hydrodynamics have been investigated 
in Ref.~\cite{Osada:2008sw} and Ref.~\cite{Biro:2011bq} respectively.  
It is found in Ref.~\cite{Biro:2011bq} that in the $q$-hydrodynamics, the main tensorial structure 
of energy momentum tensor remains same as the $d$-hydrodynamics whereas, changes appear only in 
thermodynamic quantities and EoS. Here we use this observation and derive the nonlinear wave equation 
from same tensorial structure of second order 
MIS theory, with thermodynamic quantities and equation of state from non-extensive thermodynamics. 
The fluid velocity field considered here has non-extensive nature.

Relativistic viscous hydrodynamics is a useful tool to describe the space-time evolution of the fluid 
created in relativistic heavy-ion collisions (RHICs). 
Depending on the magnitude of the expansion gradient, the order of the theory is being anticipated. 
Navier-Stokes (NS) theory deals with first-order gradients of hydrodynamic fields, usually designated as the 
first-order viscous hydrodynamics~\cite{Eckart:1940te,fluidmechanics_landau}. The solution of the NS equation is acausal and unstable. 
However, there are recently developed theories which shows that the first-order hydrodynamics can be stable as 
well as causal~\cite{Bemfica:2017wps,Bemfica:2019knx,Bemfica:2020zjp,Kovtun:2019hdm,Das:2020fnr,Das:2020gtq}.

The second-order hydrodynamics takes into account the second order gradient in hydrodynamic fields and 
the relaxation effects were introduced to attain causality and stability. Several prescriptions of the 
second-order hydrodynamics can be found in Refs.~\cite{Israel:1979wp,Hiscock:1983zz,Koide:2006ef,Romatschke:2009im,Van:2007pw,Tsumura:2009vm}. 
In this work, the second-order MIS hydrodynamics~\cite{Israel:1979wp}
is used to obtain the required evolution equations for nonlinear waves. 
For a fluid without conserved quantum number, the choice of Landau frame~\cite{fluidmechanics_landau} is very natural. However, 
Landau choice does not lead to a simple behaviour in the limit of low viscosities. But with the presence of high baryon number in a system, and with the presence of the other viscosities such as shear and bulk viscosities (which are not so low), our obvious choice of defining the fluid four-velocity is Eckart's choice of frame~\cite{Eckart:1940te}.  Eckart frame represents a local rest frame where the net charge 
dissipation is vanishing but the energy dissipation is non-vanishing. 
Although we will use Eckart frame here, one may choose the Landau frame as well. 
In Eckart frame, the energy-momentum tensor (EMT) and particle current can be written as:
\beqa
T^{\mu\nu}&=&\epsilon u^{\mu}u^{\nu}-p \Delta^{\mu\nu}+\Delta T^{\mu\nu}~,\\
J^{\mu}&=&nu^{\mu}~,
\eeqa
where, $\epsilon$ is the local energy density field, $u^{\mu}$ is the fluid four velocity, 
$p$ is the local pressure, and $\Delta T^{\mu\nu}$ 
is the correction to the EMT due to dissipation. 
 The $u^\mu$ satisfies the condition
$u^{\mu}u_{\mu}=1$, implying, $u^{\mu}\partial_{\nu}u_{\mu}=0$.
The projection operator normal to $u^{\mu}$ is defined as 
$\Delta^{\mu\nu}=g^{\mu\nu}-u^{\mu}u^{\nu}$ such that, $\Delta^{\mu\nu}u_{\nu}=0$
and $\Delta^{\mu\nu}\Delta_{\mu\nu}=3$. The symmetric traceless projection 
normal to $u^{\mu}$ is defined as, 
$\Delta^{\mu\nu}_{\alpha\beta}=
\frac{1}{2}(\Delta^{\nu}_{\alpha}\Delta^{\mu}_{\beta}+\Delta^{\mu}_{\alpha}\Delta^{\nu}_{\beta}-
\frac{2}{3}\Delta^{\mu\nu}\Delta_{\alpha\beta})~$.

The dissipative part in $T^{\mu \nu}$ can be decomposed in terms of scalar, vector and tensor as:
\beqa
\Delta T^{\mu\nu}=-\Pi \Delta^{\mu\nu}+q^{\mu}u^{\nu}+q^{\nu}u^{\mu}+\pi^{\mu\nu}~.
\eeqa
 The energy momentum tensor in Eckart frame then reads as:
 \beqa
 T^{\mu\nu}=\epsilon u^{\mu}u^{\nu}-(p+\Pi) \Delta^{\mu\nu}+q^{\mu}u^{\nu}+q^{\nu}u^{\mu}+\pi^{\mu\nu}~.
 \eeqa
In the fluid rest frame, the vector and tensor form of dissipations are considered to be 
non-existent that is,  $u_{\mu}q^{\mu}=0$, $u_{\mu}\pi^{\mu\nu}=0$. The scalar dissipation 
($\Pi$) corresponds to the non-equilibrium pressure perturbation, which can not be made to relate to 
energy density using the EoS. It is related to volume expansion of the fluid, triggers small 
non-equilibrium perturbation, which tends to  bring 
a new equilibrium state through the dissipative correction. 
The vector $q^{\mu}$ and the $\pi^{\mu \nu}$ 
correspond to dissipative energy flow and the shear stress tensor respectively. 
The relations between energy density, 
pressure and the dissipative fluxes and EMT are given by the following relations:
\beqa
u_{\mu}T^{\mu\nu}u_{\nu}&=&\epsilon,\,\,\,\, q_{\alpha}=u_{\mu}T^{\mu\nu}\Delta _{\nu\alpha}=u_{\mu}\Delta T^{\mu\nu}\Delta _{\nu\alpha}, \nn\\
p+\Pi&=&-\frac{1}{3}\Delta_{\mu\nu}T^{\mu\nu},\,\,\,\, \Pi=-\frac{1}{3}\Delta_{\mu\nu}\Delta T^{\mu\nu},\,\,\,\, u_{\mu}\Delta T^{\mu\nu}=q^{\nu}~.
\eeqa
The equation of motion are obtained from the conservation of EMT and the net charge density (here the net baryon number) density as:
\beqa
\partial_{\mu}T^{\mu\nu}&=&0~,
\label{eq6}
\eeqa
\beqa
\partial_{\mu} J^{\mu}&=&0~.
\label{eq7}
\eeqa
The dissipative fluxes~\cite{Israel:1979wp} are given by,
\beqa
\label{eq8}
\Pi&=& -\frac{1}{3}\zeta\Big[\pd_{\mu}u^{\mu}+\beta_{0}D\Pi-\tilde{\alpha_{0}}\pd_{\mu}q^{\mu}\Big]~, \\
\label{eq9}
\pi^{\lambda \mu}&=&-2\eta \Delta^{\lambda\mu\alpha\beta}\Big[\partial_{\alpha}u_{\beta}+\beta_{2}D\pi_{\alpha\beta}-\tilde{\alpha_{1}}\partial_{\alpha}q_{\beta}\Big]~,\\
\label{eq10}
q^{\lambda}&=&-\kappa T\Delta^{\lambda\mu} \Big[\frac{1}{T}\partial_\mu T +Du_{\mu}+\tilde{\beta_1} D{q_\mu}-\tilde{\alpha_0}\partial_\mu \Pi -\tilde{\alpha_1}\pd_{\nu}\pi ^{\nu}_{\mu}~. \Big] 
\eeqa
Here, $\eta$, $\zeta $, $\kappa$ are the coefficient of shear viscosity, bulk viscosity and thermal conductivity respectively. $D\equiv u^\mu\partial_\mu$, 
is the co-moving derivative and in the local rest frame (LRF) it represents  
the time derivative, $D\Pi =\dot{\Pi}$. The quantities,
$\beta _0,\tilde{\beta_1}, \beta_2$ are relaxation coefficients, 
$\tilde{\alpha_0}$ and $\tilde{\alpha_1}$ are coupling coefficients. 
The $\beta _0,\tilde{\beta_1}$ and $\beta_2$ are also related to the relaxation time scale as~\cite{Muronga:2003ta,Muronga:2001zk}:
\begin{equation}
\tau_{\Pi}=\zeta \beta_0, \,\,\,\,\tau_q=\kappa T\beta_1,\,\,\,\, \tau _{\pi}=2\eta \beta_2
\label{eq12}
\end{equation}
And the coupling coefficients are related to the relaxation lengths, which couple to heat flux, and bulk pressure $(l_{q\Pi}, l_{\Pi q})$, the heat flux, and shear tensor $(l_{q\pi}, l_{\pi q})$ by the following relation
\begin{equation}
l_{\Pi q}=\zeta \alpha_0,\,\,\,\, l_{q\Pi}=\kappa T \alpha _0, \,\,\,\,l_{q\pi}=\kappa T\alpha_1,\,\,\,\, 
l_{\pi q}=2\eta \alpha_1~.
\label{eq13} 
\end{equation}  
The  expressions for relaxation and coupling coefficients are given in Appendix-\ref{appendixA}.

\subsection{1-D flow equations}
In this section the relevant equations are derived to study the propagation of nonlinear perturbation in a 
fluid in $(1+1)$ dimension. 
The evolution equations of energy density and fluid velocity are obtained from  
the Eqs.\eqref{eq6} and \eqref{eq7} by taking projection along the directions parallel and perpendicular to $u^{\mu}$ as:
\beqa
\label{eq17}
D\epsilon=-(\epsilon+p)\theta+(\partial_{\mu} u_{\nu})\Delta T^{\nu\mu}-(\partial_{\mu} q^{\mu})~,\\
\label{eq18}
(\epsilon+p)D u^{\alpha}=\Delta^{\alpha\mu}\partial_{\mu}-\Delta^{\alpha}_{\nu}\partial_{\mu} \Delta T^{\nu\mu}~,
\eeqa
where, $u^{\mu}=(\gamma,\gamma v, 0,0),\,\,\,\, \gamma=1/\sqrt{1-v^{2}}$ and
\beqa
\theta&=&\partial_{\mu}u^{\mu}=v\gamma^3\frac{\partial v}{\partial t}+\gamma^3\frac{\partial v}{\partial x}~,\\
\partial_{\mu}q^{\mu}&=&q^{x}\frac{\partial v}{\partial t}+v\frac{\partial q^{x}}{\partial t}+\frac{\partial q^{x}}{\partial x}~.
\eeqa
Here, 
\beqa
u_{\mu}q^{\nu}=0,\,\,\, u_{\mu}\pi^{\mu\nu}=0,\,\,\, g_{\mu\nu}\pi^{\mu\nu}=0,\,\,\, \pi^{xy}=\pi^{zx}=\pi^{yz}=0,\,\,\, \pi^{yy}=\pi^{zz}=\pi^{T}
\eeqa
which leads to, 
\beqa
\pi^{00}=v^2\pi^{xx},\,\,\, \pi^{xt}=\pi^{tx}=
v^2\pi^{xx},\,\,\, \pi^{T}=\pi^{tx}=\frac{(
v^2-1)}{2}\pi^{xx}~,
\eeqa
and 
\beqa
q^{t}=vq^{x}, \,\,\,q^{y}=q^{z}=
q^{T}~.
\eeqa
Now from the energy-momentum conservation equations {\it{i.e} } from Eqs.\eqref{eq17}, and \eqref{eq18}, we get,
\beqa
\label{enc}
\gamma v\Big(\frac{\partial \epsilon}{\partial t}+\frac{\partial \epsilon}{\partial x}\Big)=&& \Big[ (h+\Pi)v\gamma^3+v \gamma \pi^{xx}+q^{x}\}\frac{\partial v}{\partial t} \Big]\nn\\
&&+\Big[(h+\Pi)\gamma^3+\gamma \pi^{xx}\Big]
\frac{\partial v}{\partial x}+v\frac{\partial q^{x}}{\partial t}+\frac{\partial q^{x}}{\partial x},
\eeqa
\beqa
\label{mc}
&&\Big[ (h+\Pi)v\gamma^4+v \gamma^2(1-2v^2)v \pi^{xx}+\gamma^3(1+v^2)v
q^{x}\Big]\frac{\partial v}{\partial t}+\Big[ (h+\Pi)v\gamma^4-\gamma^2 v^2\pi^{xx}\nn\\
&&+\gamma v(1+2\gamma^2)q^{x}\Big]
\frac{\partial v}{\partial x}
=-v^2\Big[\gamma^2\frac{\partial p_{\zeta}}{\partial t}+\frac{\gamma^2 }{v}\frac{\partial p_{\zeta}}{\partial x}
+\frac{\partial \pi^{xx}}{\partial t}+ \frac{1}{v}\frac{\partial \pi^{xx}}{\partial x}
+\frac{ \gamma}{v}\frac{\partial q^{x}}{\partial t}- \gamma\frac{\partial q^{x}}{\partial x}\Big]
\eeqa
where $h=\epsilon+p$ is the enthalpy density. We define $p+\Pi=p_{\zeta}$. 
The equation governing the net baryon number conservation reads,
\beqa
\label{nc}
\frac{\partial n}{\partial t}+ v\frac{\partial n}{\partial x}
= -n \gamma^2\Big[v\frac{\partial v}{\partial t}+\frac{\partial v}{\partial x}\Big]~,
\eeqa
The other three equations originating from dissipative fluxes are given by the Eqs.\eqref{eq8}, \eqref{eq9}, and \eqref{eq10}, which 
can be written as:
\beqa
\label{fl1}
\Pi+\zeta \beta_{0}\Big[\gamma \frac{\partial \Pi}{\partial t}+v\frac{\partial \Pi}{\partial x}\Big]&=&-\zeta\Big[\{v\gamma^3 -\alpha_{0} q^{x}\}\frac{\partial v}{\partial t}+\{ \gamma^3\}
\frac{\partial v}{\partial x}-\alpha_{0}v\frac{\partial q^{x}}{\partial t}-\alpha_{0}\frac{\partial q^{x}}{\partial x}\Big]~,\\ \nn\\
\label{fl2}
 q^{x}+\kappa T \beta_{1}\Big[\gamma \frac{\partial q^{x}}{\partial t}+v\frac{\partial q^{x}}{\partial x}\Big]&=& \kappa T\Big[
\{-\gamma^4+ \beta_{1}\gamma^3 vq^{x}+\alpha_{1}(2-\gamma^2) \pi^{xx}
 \}\frac{\partial v}{\partial t}+\{-v\gamma^4\nn\\
 &&+ \beta_{1}\gamma^3 v^2 q^{x}- \alpha_{1}\gamma^2 v \pi^{xx}
 \}
\frac{\partial v}{\partial x}
+\alpha_{1}v \frac{\partial \pi^{xx}}{\partial t}+\alpha_{1} \frac{\partial \pi^{xx}}{\partial x}\nn\\
&&-\alpha_{0}v\gamma^2 \frac{\partial \Pi}{\partial t}-\alpha_{0}\gamma^2 \frac{\partial \Pi}{\partial x}
+\frac{1}{T}v\gamma^2 \frac{\partial T}{\partial t}+\frac{1}{T}\gamma^2 \frac{\partial T}{\partial x}\Big]~,\\ \nn\\
\label{fl3}
\pi^{xx}+\frac{4}{3}\eta \beta_{2}\Big[\gamma \frac{\partial\pi^{xx}}{\partial t}+v\frac{\partial \pi^{xx}}{\partial x}\Big]&=& \frac{4}{3}\eta\Big[
\{\gamma^5+ \alpha_{1}\gamma^4 v q^{x}+ 2\beta_{2}\gamma^3 \pi^{xx}
 \}\frac{\partial v}{\partial t}+\{\gamma^5+ \alpha_{1}\gamma^4 v q^{x}\nn\\
 &&+ 2\beta_{2}\gamma^3v^2 \pi^{xx}
 \}
\frac{\partial v}{\partial x}-\alpha_{1}\gamma^2 v\frac{\partial q^{x}}{\partial t}-\alpha_{1}\gamma^2 \frac{\partial q^{x}}{\partial x}]~.
\eeqa

\subsection{Derivation of nonlinear wave equations}
The above hydrodynamic equations from MIS causal theory is used to derive the nonlinear equations governing
the motion of the perturbations in the fluid. 
We have adapted Reductive Perturbative Method (RPM)~\cite{PhysRevLett.17.996,Davidson:plasma_book,Leblond_2008, 1995patt.sol..9003K}.  To proceed further
we need to define `stretched co-ordinates' as,
\beqa
X=\frac{\sigma^{1/2}}{L}(x-c_s t), \text{and} \hspace{0.4cm} Y=\frac{\sigma^{3/2}}{L}(c_s t)~,
\eeqa
where, $L$ is the characteristic length and $c_{s}$ is the speed of sound. Therefore, from the above equations we get
\beqa
\label{ct}
\frac{\partial}{\partial x}=\frac{\sigma^{1/2}}{L}\frac{\partial}{\partial X}, \text{and} \hspace{0.4cm} 
\frac{\partial}{\partial t}=-c_{s}\frac{\sigma^{1/2}}{L}\frac{\partial}{\partial X}+c_{s}\frac{\sigma^{3/2}}{L}\frac{\partial}{\partial Y}~,
\eeqa
where $\sigma$ is the expansion parameter. The coordinate $X$ is measured from the frame of propagating 
sound waves, whereas the $Y$ represents fast moving coordinate. 
The RPM technique is devised to preserve the structural form of the parent equation.  
Now we perform the simultaneous series expansion of hydrodynamic quantities in  powers of
$\sigma$ to get,
\beqa
\hat{\epsilon}&=&\frac{\epsilon}{\epsilon_0}=1+\sigma \epsilon_1+\sigma^2 \epsilon_2+\sigma^3 \epsilon_3+...,\,\,\,
\hat{p}=\frac{p}{p_0}=1+\sigma p_1+\sigma^2 p_2+\sigma^3 p_3+...,\nn\\
\hat{v}&=&\frac{v}{c_s}=\sigma v_1+\sigma^2 v_2+\sigma^3 v_3+...,\,\,\,\,\,\,\,\,\,\,\,\,
\hat{\Pi}=\frac{\Pi}{p_0}=\sigma \Pi_1+\sigma^2 \Pi_2+\sigma^3 \Pi_3+...,\nn\\
\hat{q}^x&=&\frac{q^x}{\epsilon_0}=\sigma q^x_1+\sigma^2 q^x_2+\sigma^3 q^x_3+...,\,\,\,\,\,\,\,
\hat{\pi}^{xx}=\frac{\pi^{xx}}{p_0}=\sigma \pi^{xx}_1+\sigma^2 \pi^{xx}_2+\sigma^3 \pi^{xx}_3+...
\label{epshat}
\eeqa
By collecting terms with different order of $\sigma$ from 
the perturbation series different types of equation like 
Breaking wave equation, Burgers equation or KdV equation, etc can be
obtained~\cite{Fogaca:2009wf,Lick1990,Bhattacharyya:2020sua}. 

We rewrite the equations of motions (Eqs.~\ref{enc},~\ref{mc},~\ref{nc},~\ref{fl1},~\ref{fl2},~\ref{fl3}) using Eqs.~\ref{ct},~\ref{epshat} and collect terms with order upto $\sigma^3$ 
and finally revert from $(X,Y) \to (t,x)$ to get the following equations 
for the perturbation in the energy density ($\hat{\epsilon}$) as:
\beqa
\frac{\pd \hat{\epsilon_{1}}}{\pd t}
+\left[1+(1-c_{s}^2) \frac{\epsilon_{0}}{\epsilon_{0}+p_{0}}\hat{\epsilon_{1}}\right]c_{s}\frac{\pd \hat{\epsilon_{1}}}{\pd x}- \left[\frac{1}{2(\epsilon_{0}+p_{0})}(\zeta +\frac{4}{3}\eta)\right]
\frac{\pd^{2} \hat{\epsilon_{1}}}{\pd x^{2}}=0,
\label{nlw1}
\eeqa
and
\beqa
\frac{\pd \hat{\epsilon_{2}}}{\pd t}+\mathcal{S}_{1}\frac{\pd \hat{\epsilon_{2}}}{\pd x}+\mathcal{S}_{2}\frac{\pd \hat{\epsilon_{1}}}{\pd x}+\mathcal{S}_{3} \frac{\pd^{2} \hat{\epsilon_{1}}}{\pd x^{2}}+\mathcal{S}_{4}\frac{\pd^{3} \hat{\epsilon_{1}}}{\pd x^{3}}+\mathcal{S}_{5}\frac{\pd^{2} \hat{\epsilon_{2}}}{\pd x^{2}}=0~.
\label{nlw2}
\eeqa
where $\hat{\epsilon_{i}}= \sigma^i \epsilon_i$ for $i=1,2,.....$.
The coefficients $\mathcal{S}_i$'s for $i=1$ to $5$ are given in Appendix-\ref{appendixB}.
Eqs.~\eqref{nlw1} and \eqref{nlw2} have been solved to investigate the fate of the 
nonlinear waves propagating through a relativistic viscous and non-extensive
fluid. The relevant EoS and the initial conditions required to solve
these equations are discussed in section III and IV respectively.

\section{Tsallis MIT bag equation of state}
\label{sec3}
We assume that the quarks and gluons in the viscous plasma medium follow the quantum Tsallis 
Fermionic ($F$) and Bosonic ($B$) single particle distributions given by~\cite{Rahaman:2021xqv},
\bea
f_{Fq}=\frac{1}{\left[1+(q-1)\frac{E_p-\mu}{T}\right]^{\frac{q}{q-1}}+1}\nn\\
f_{Bq}=\frac{1}{\left[1+(q-1)\frac{E_p-\mu}{T}\right]^{\frac{q}{q-1}}-1}
\label{TsFDBE},
\eea
where, $E_p=\sqrt{p^2+m^2}$ is the single particle energy of a particle of mass $m$, 
$\mu$ is the (baryonic) chemical potential (assumed to be zero in the present work), 
$q$ is the Tsallis (non-extensitivity) parameter and $T$ is the Tsallis temperature. 
The $\epsilon$ and $p$ of a system of quarks and gluons in MIT bag model 
~\cite{PhysRevD.9.3471} is given by,
\bea
\epsilon_{\mathrm{}} &=& \mathcal{B}+\epsilon_{B}+2\epsilon_F, \label{bagepsilon}\\
P_{\mathrm{}} &=& -\mathcal{B}+P_{B}+2P_F, \label{bagP}
\eea

where $\mathcal{B}$ is the bag constant ($\mathcal{B}^{1/4}= 200$ MeV). 
The $\epsilon_{i}$ and $P_{i}$ are given by, 
\bea
\epsilon_i = g \int \frac{d^3p}{(2\pi)^3}E_p f_{iq},~ \ \ \ \ 
P_i = g \int \frac{d^3p}{(2\pi)^3}\frac{p^2}{3E_p} f_{iq},
\eea
where $g$ is the degeneracy factor and $i=B, F$.

The factor 2 in front of the Fermionic contributions  in Eqs.~\eqref{bagepsilon} and ~\eqref{bagP} 
takes into account both quarks and anti-quarks. 
The net baryonic number density, $n$ is assumed to be zero here. The Fermionic thermodynamic variables
have been evaluated up to $\mathcal{O}(m^2T^2)$.  
It is assumed that the quark-gluon plasma consists of the up quark, down quark and gluons. 
It has been verified that for a wide range of the values of $q$ and $T$ relevant for 
phenomenological studies of high-energy collisions, the $\mathcal{O}(m^2T^2)$ approximated 
results overlap with the exact results when the mass is small ($\sim$10 MeV). 
We write down the expressions 
for the energy density and pressure in the non-extensive bag 
model~\cite{Lavagno:2010xu,Cardoso:2017pbu} as:

\bea
\epsilon_{\mathrm{}} &=& \mathcal{B} + \epsilon_{B,1}T^4 + 2 (\epsilon_{F,1}T^4+\epsilon_{F,2}m^2T^2)
				     = \mathcal{B} + \epsilon_{\alpha}T^4 + \epsilon_{\beta}m^2T^2 \label{ebagshort}~, \\ \nn\\
P_{\mathrm{}}         &=&  -\mathcal{B} + P_{B,1}T^4 + 2 (P_{F,1}T^4+P_{F,2}m^2T^2)
				      =  -\mathcal{B} + P_{\alpha}T^4 + P_{\beta}m^2T^2. \label{Pbagshort}
\eea
where,
\bea
\epsilon_{\alpha} &=& \epsilon_{B,1} + 2\epsilon_{F,1}, ~P_{\alpha} = P_{B,1} + 2P_{F,1}, \nn\\
~\epsilon_{\beta} &=& 2\epsilon_{F,2},~P_{\beta} = 2P_{F,2}. 
\eea
The expressions for $\epsilon_{i,\ell}$ and $P_{i,\ell}$ ($i=F, B$; $\ell=1,2$) are given in the Appendix-\ref{appendixC}.

The expressions for pressure and energy density obtained in this section are used to solve
the Eqs.~(\ref{nlw1}) and~(\ref{nlw2}), governing the evolution of the nonlinear perturbations

\section{Results and discussion}
\label{sec4}
We present here the results on the fate of the nonlinear perturbations in 
the QGP fluid at zero baryonic chemical potential which leads to the 
vanishing thermal conductivity of the medium. 
We solve the equations describing the evolution of $\hat{\epsilon_i}$ 
for the following initial profile of the perturbations:
\beqa
\hat{\epsilon}_1&=&A_{1}\big[\mathrm{sech} \frac{(x-x_{0})}{B_{1}}\big]^{2} \nn\\
\hat{\epsilon}_2&=&A_{2}\big[\mathrm{sech} \frac{(x-x_{0})}{B_{2}}\big]^{2}
\label{inprof}
\eeqa
where, we have taken $x_{0}=10$ fm. The Eqs.\eqref{nlw1}, and \eqref{nlw2} 
will reduce to KdV type equations for small dissipations.
We assume that the initial profile of the perturbations are  $\mathrm{sech}$ function 
as the solitonic solution of the KdV equation involves  $\mathrm{sech}$  functional dependence.

\begin{figure}[h]
\centering
\includegraphics[width=7.5cm]{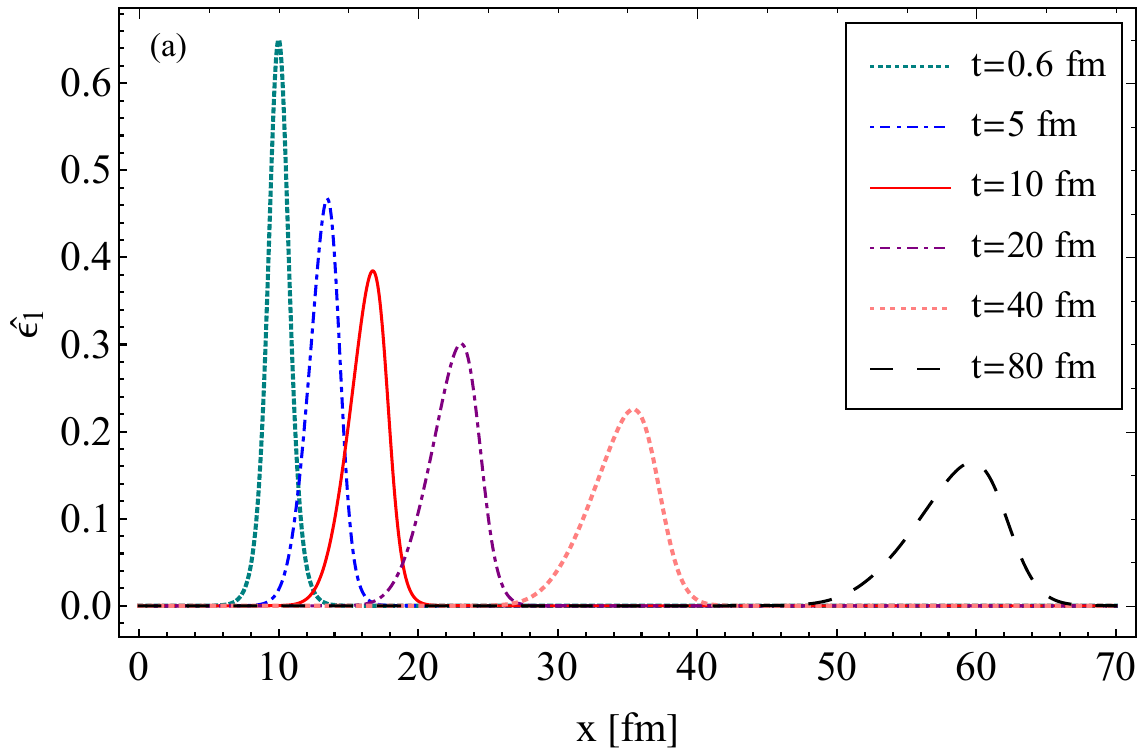}	
\includegraphics[width=7.7cm]{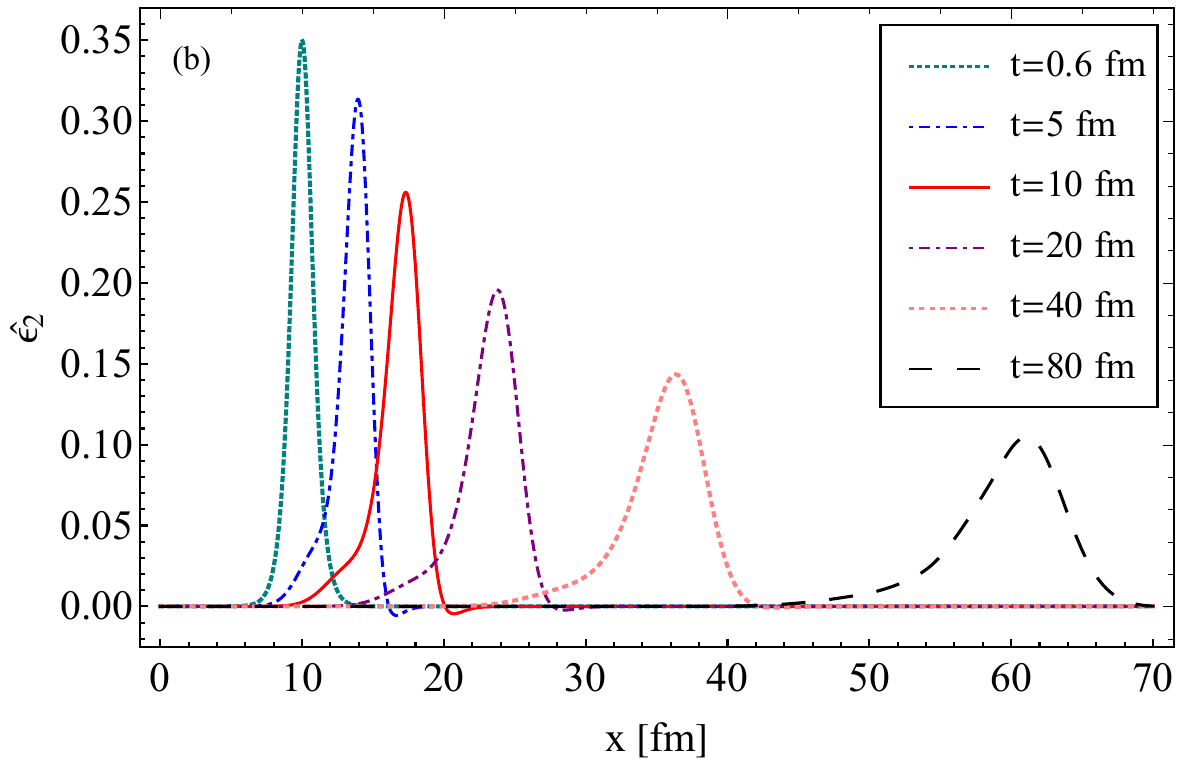}\\
\includegraphics[width=9.2cm]{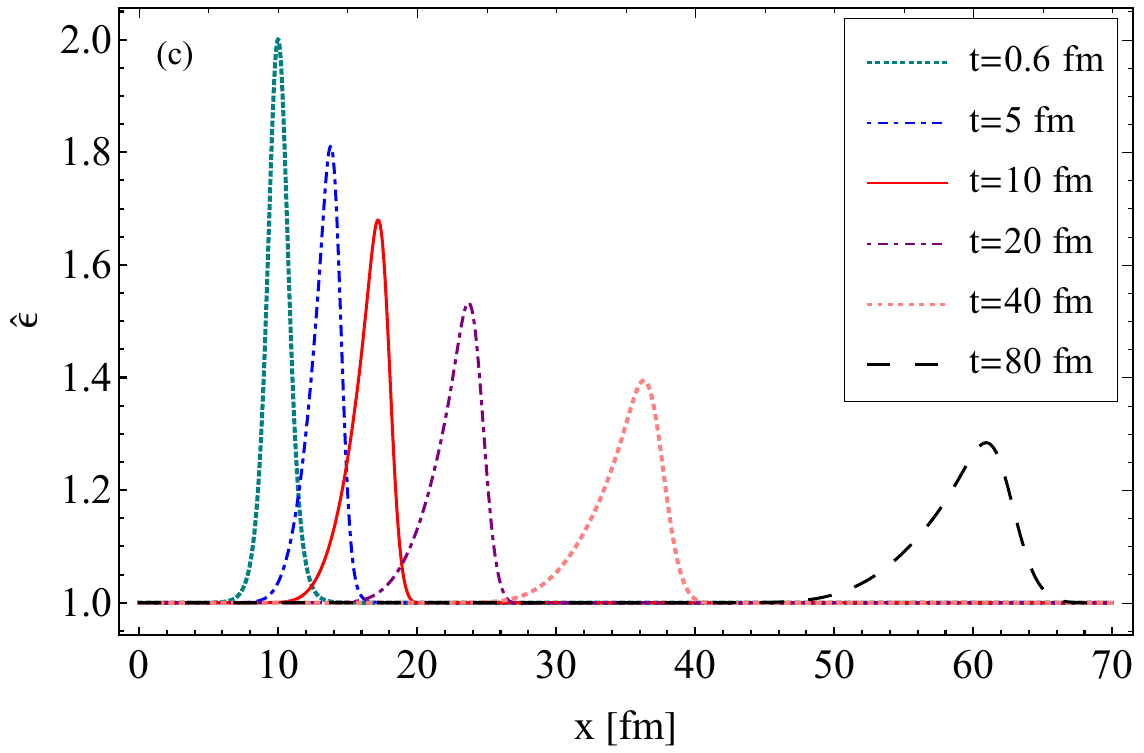}	
\caption{(Color online) (a) $\hat{\epsilon}_1$ as a function of $x$ at different time for 
$A_{1}=0.65$ and $B_{1}=1$ fm. 
(b) $\hat{\epsilon}_2$ as a function of $x$ at different time for $A_{2}=0.35$ and $B_{2}=1$ fm. 
(c) $\hat{\epsilon}=1+\hat{\epsilon}_1+\hat{\epsilon}_2$ as a function of $x$ at different time 
for $A_{1}=0.65, B_{1}=1$ fm, $A_{2}=0.35$ $B_{2}=1$ fm. The background temperature,
$T=200$ MeV and $q=1.08$ here.}
\label{fig1}
\end{figure}

\begin{figure}[h]
\centering
\includegraphics[width=9.5cm]{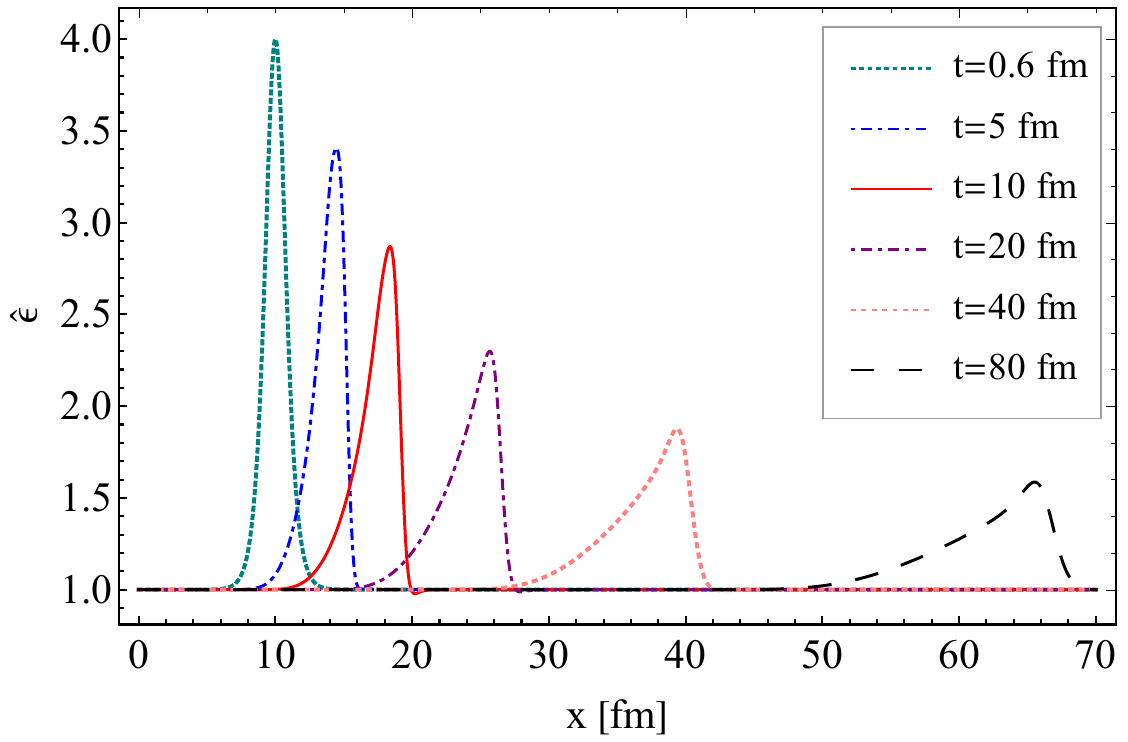}	
\caption{(Color online) $\hat{\epsilon}$ as a function of $x$ at different time for $T= 200 MeV$,
$q= 1.08$, $A_{1}=1.65, B_{1}=1, A_{2}=1.35, B_{2}=1$ (parameters are same as in Fig.1(c) but
with higher amplitude). We observe that with larger amplitude the speed is larger.}
\label{fig2}
\end{figure}

\begin{figure}[h]
\centering
\includegraphics[width=7.5cm]{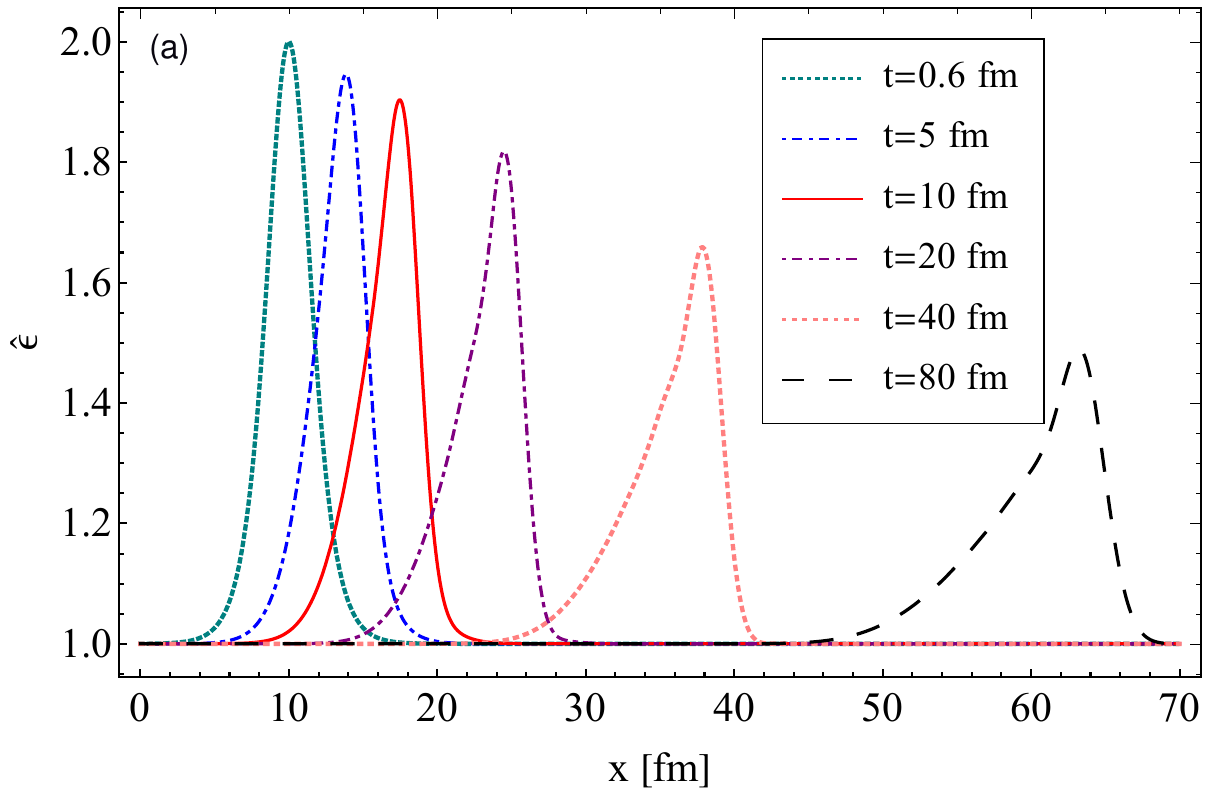}	
\includegraphics[width=7.5cm]{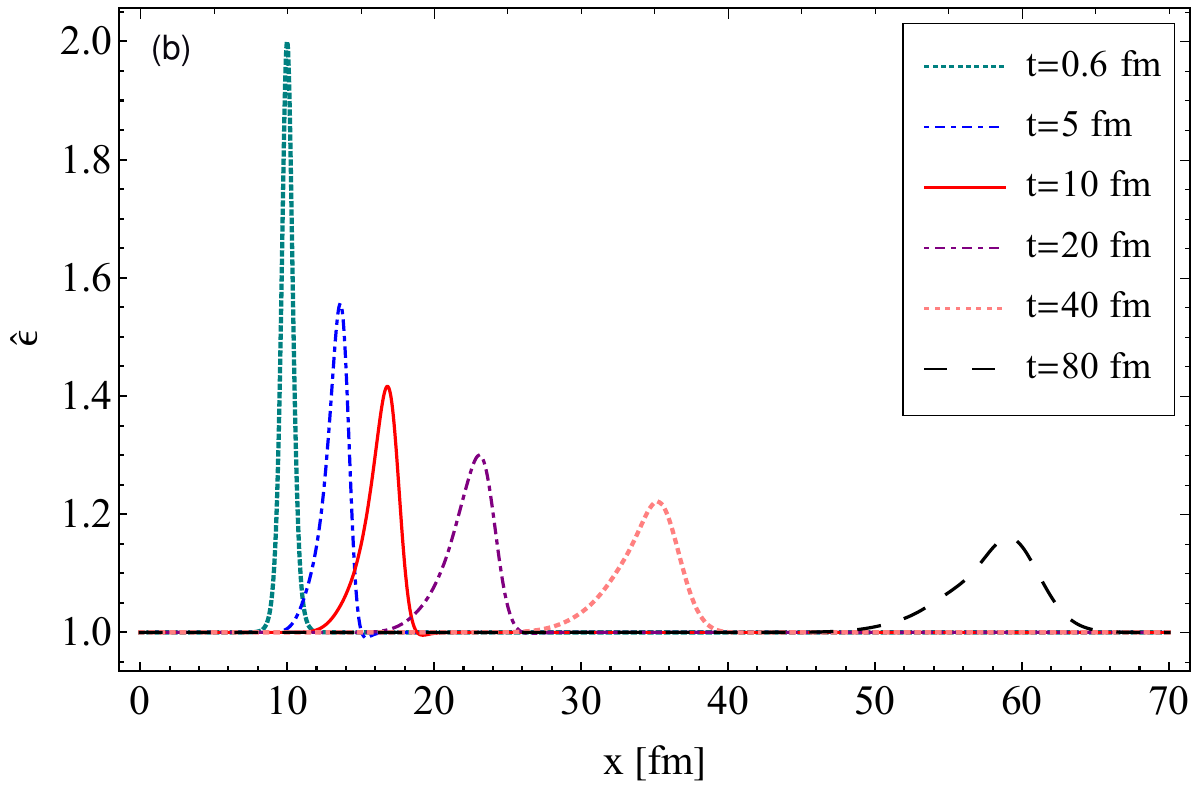}	
\caption{(Color online) (a) $\hat{\epsilon}$ as a function of $x$ at different time for $T= 200 MeV$, $q= 1.08$ 
$A_{1}=0.65, B_{1}=2, A_{2}=0.35$  and $B_{2}=2$ (parameters are same as in Fig.1 (c)) but with higher width). 
We observe that the wave propagates faster with lower dissipation as the width increases.
(b) Same as Fig.~\ref{fig1}(c) with $A_{1}=0.65, B_{1}=0.5, A_{2}=0.35,$ and $B_{2}=0.5$. 
In this case larger dissipation and smaller speed of propagation is observed.}
\label{fig3}
\end{figure}

Firstly, we investigate the effects of non-extensivity only through the EoS, keeping  
the value of the shear viscosity at the KSS bound~\cite{Kovtun:2004de},
$\eta/s=1/4\pi$. The bulk viscosity is taken as $\zeta/s= 1/4\pi$ and $\kappa= 0$. 
The value of $T$ and $q$ are taken as $T=200 MeV$ and $q=1.08$. 
Fig.~\ref{fig1}(a) and Fig.~\ref{fig1}(b) show the time evolution of the spatial 
shape of the first-order $(\hat{\epsilon_1})$ and the second-order $(\hat{\epsilon_2})$ perturbations respectively. 
The initial profile is given by the Eq.~\eqref{inprof}. In Fig.~\ref{fig1}(a), the initial magnitude $(A)$ and width $(B)$ 
of the first-order perturbation are taken as $A=0.65$, $B=1fm$, 
whereas in Fig.~\ref{fig1}(b), the initial magnitude of 
the second-order perturbation is taken as $A=0.35$ and width is kept  
at $B=1fm$. 
Comparison of results depicted in Fig.~\ref{fig1}(a) and Fig.~\ref{fig1}(b) 
indicate that the dissipation 
of the second-order perturbation is less as compared to first order. 
This can be understood from the Eq.~\eqref{nlw2}, 
which indicates that $\hat{\epsilon_2}$ gets contributions from $\hat{\epsilon_1}$ 
which contains terms with third-order derivatives in space, 
known as dispersive term, responsible for preserving the shape of the
propagating nonlinear wave. 
Fig.~\ref{fig1}(c) shows the evolution of the energy density up to second order scaled by 
the background energy density. 
Though the QGP system produced in heavy-ion collisions has a lifetime of the order of $10-15fm$, 
we have considered the time evolution up to $80fm$ to 
have a better understanding of the late time behaviour. 
It is evident from the results that with time, amplitude decreases and width increases 
indicating towards the 
possibility of complete dissipation, unlike the case studied 
earlier in Ref.~\cite{Bhattacharyya:2020sua}. 
It may be noted that up to $15 fm$ the dissipation of the nonlinear wave is not very drastic. 
The change in amplitude is faster in the beginning than later times.
At later time the smaller propagation speed is also noticed.
The slowing down of the dissipation can be understood from a comparison of the results displayed in  
Fig.~\ref{fig1}(c), Fig.~\ref{fig2}, Fig.~\ref{fig3}(a), and Fig.~\ref{fig3}(b). 
In Fig.~\ref{fig1}(c) and Fig.~\ref{fig2}, the amplitudes of the 
initial perturbations are taken to be different with the same value of the 
width, whereas in Fig.~\ref{fig1}(c), the amplitude is taken as $A=2$ and 
in Fig.~\ref{fig2} it is considered to be $A=4$.
It is observed that with larger amplitude, the speed of propagation is larger, 
which is clearly seen 
in Fig.~\ref {fig2} (the result with $t=80$ fm provides better visibility). 
This is due to amplitude dependent propagation of 
nonlinear waves, a well known feature of the nonlinear propagation. 
The nonlinear waves become slower as the amplitude reduces due to dissipation. 


In Fig.~\ref{fig3}(a) and Fig.~\ref{fig3}(b), we compare the results for 
the widths taken as $B=2 fm$ and $B=0.5 fm$ respectively with the same 
value of the amplitude ($A=2$). It is observed  that the increase in width of the initial perturbation
causes less dissipation with the progression of time for the same background fluid conditions. 
The propagation speed is found to be greater for perturbation with larger width. This is because, 
the perturbation with larger width dissipates less, and thus due to amplitude dependent speed of propagation, 
it propagates faster in space. Comparison of results presented in Fig.~\ref{fig2} and  Fig.~\ref{fig3} 
indicates that the increase in amplitude or in width  reduces the damping but increases the speed of the peak position.

\begin{figure}[h]
\centering
\includegraphics[width=9.5cm]{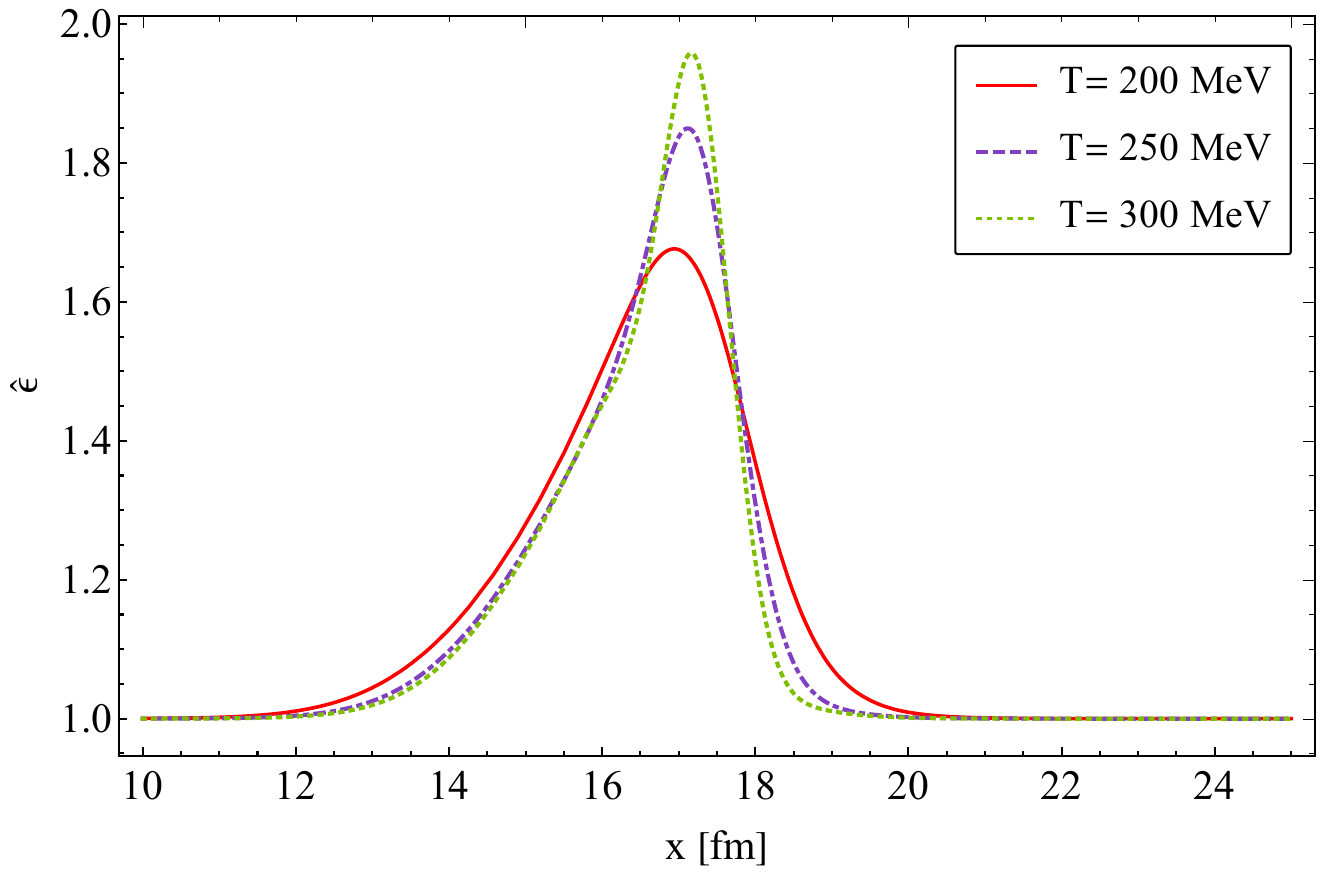}	
\caption{(Color online) $\hat{\epsilon}$ as a function of $x$ at time $t= 10$ fm for different $T$. 
Other parameters are $q= 1.08$, $A_{1}=0.65, B_{1}=1, A_{2}=0.35, B_{2}=1$. 
We observe smaller dissipation at larger temperature.
}
\label{fig4}
\end{figure}

Now we move to discuss the role of the background of the fluid, inscribed through 
$q$ and $T$, on the perturbations. We take the same values of $A_1$, $A_2$, $B_1$ and $B_2$ as in
Fig.\ref{fig1}(c) to study the fate of the perturbations after $t=10$ fm for three different values of the
background temperatures and fixed value of $q=1.08$. The result is displayed in Fig.~\ref{fig4}.
We find that the dissipation decreases with increase in the background temperature but 
the perturbation in the front becomes steep.
However, for ideal background, the magnitude as well as  the tilting 
remain unchanged with the variation of the background temperature. 
In the present work we consider the background energy density as a function of temperature
in contrast to previous work~\cite{Bhattacharyya:2020sua} on ideal 
hydrodynamic background at fixed background energy density with varying temperature.

\begin{figure}[h]
\centering
\includegraphics[width=9.5cm]{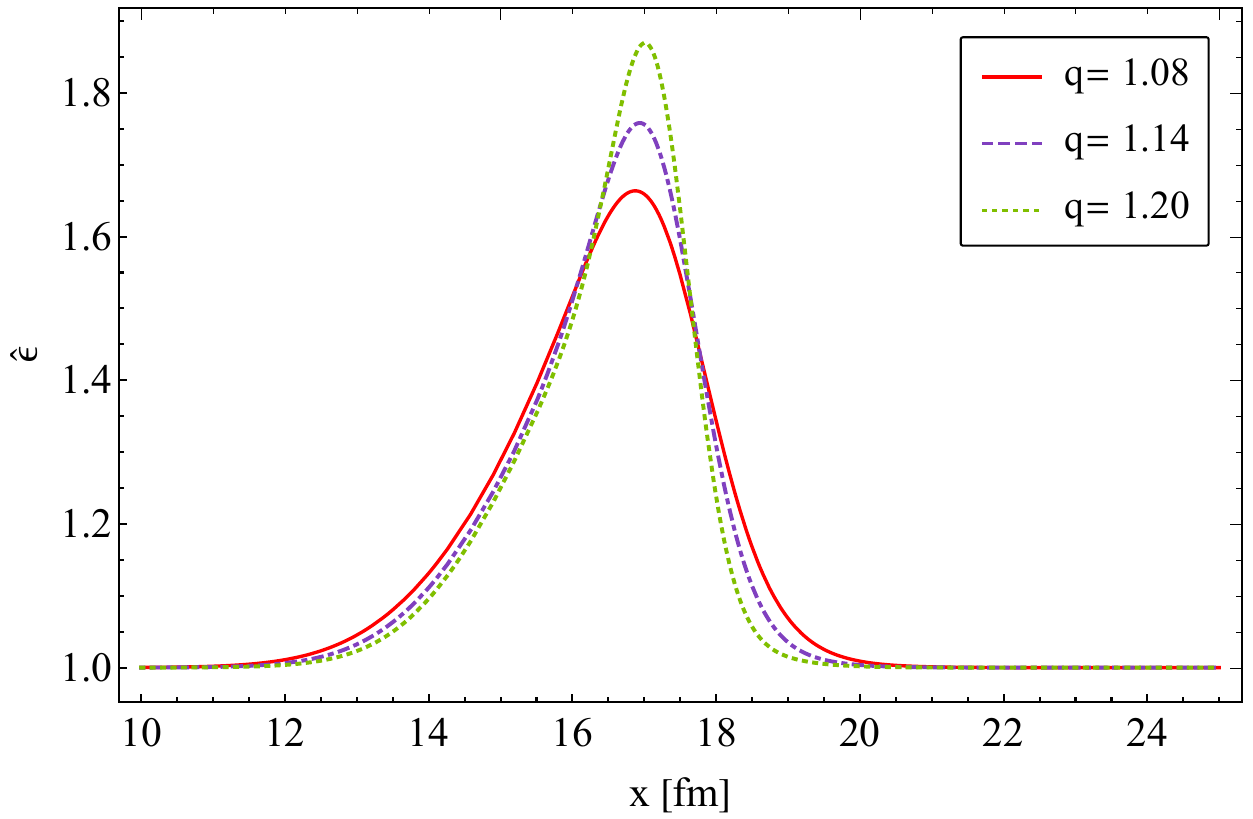}	
\caption{(Color online) $\hat{\epsilon}$ as a function of $x$ at time $t= 10$ fm and $T= 200$ MeV for
different values of $q$. The amplitudes and the widths of the perturbations are
defined by Eq.~\eqref{inprof} with $A_{1}=0.65, B_{1}=1, A_{2}=0.35, B_{2}=1$. 
Smaller dissipation and tilting along $x$ axis have been seen for larger $q$. 
}
\label{fig5}
\end{figure}
\begin{figure}[h]
\centering
\includegraphics[width=9.5cm]{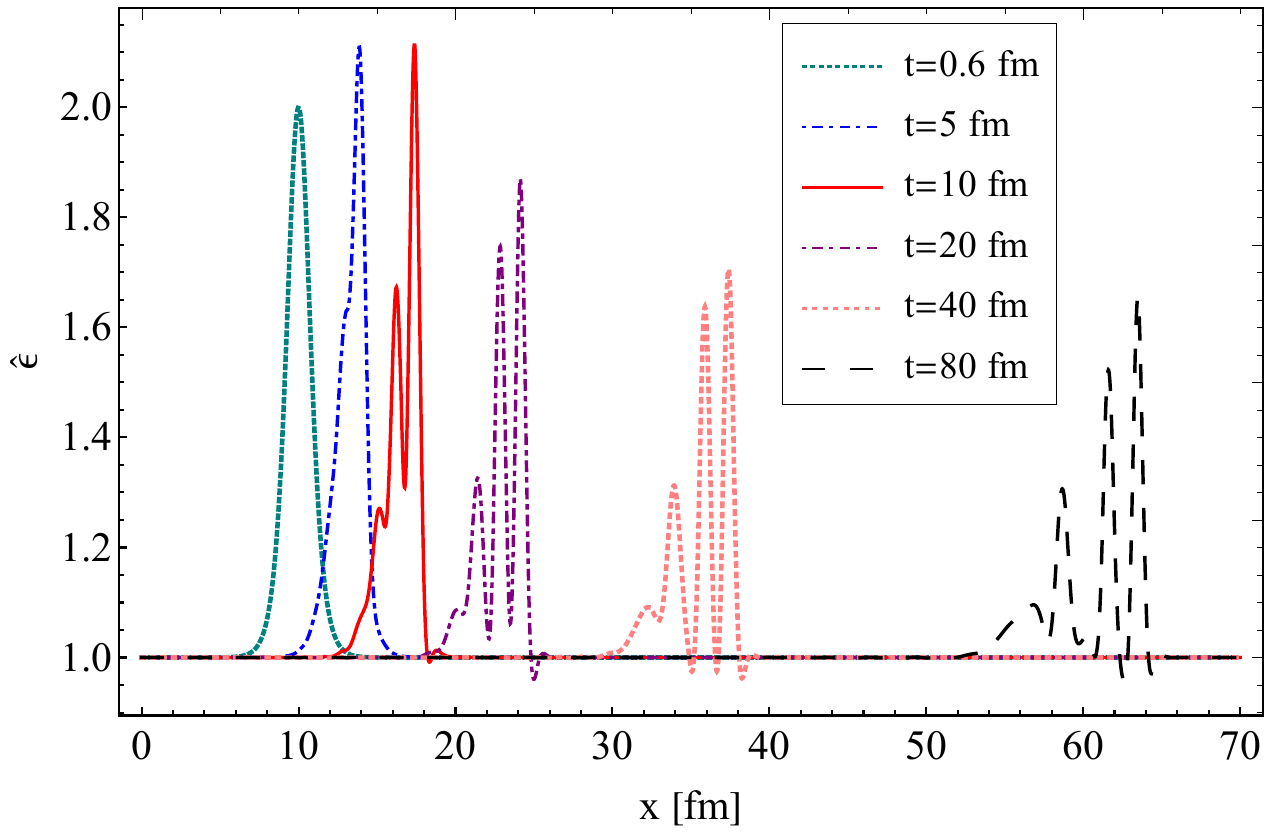}	
\caption{(Color online) $\hat{\epsilon}$ as a function of $x$ for different time 
for ideal fluid ($\eta= \zeta = 0$). We take $T=200$ MeV,
$q=1.08$, $A_{1}=0.65, B_{1}=1, A_{2}=0.35$ and $B_{2}=1$. 
We observe that the waves tend to loose their localization with time.}
\label{fig6}
\end{figure}

\begin{figure}[h]
\centering
\includegraphics[width=9.5cm]{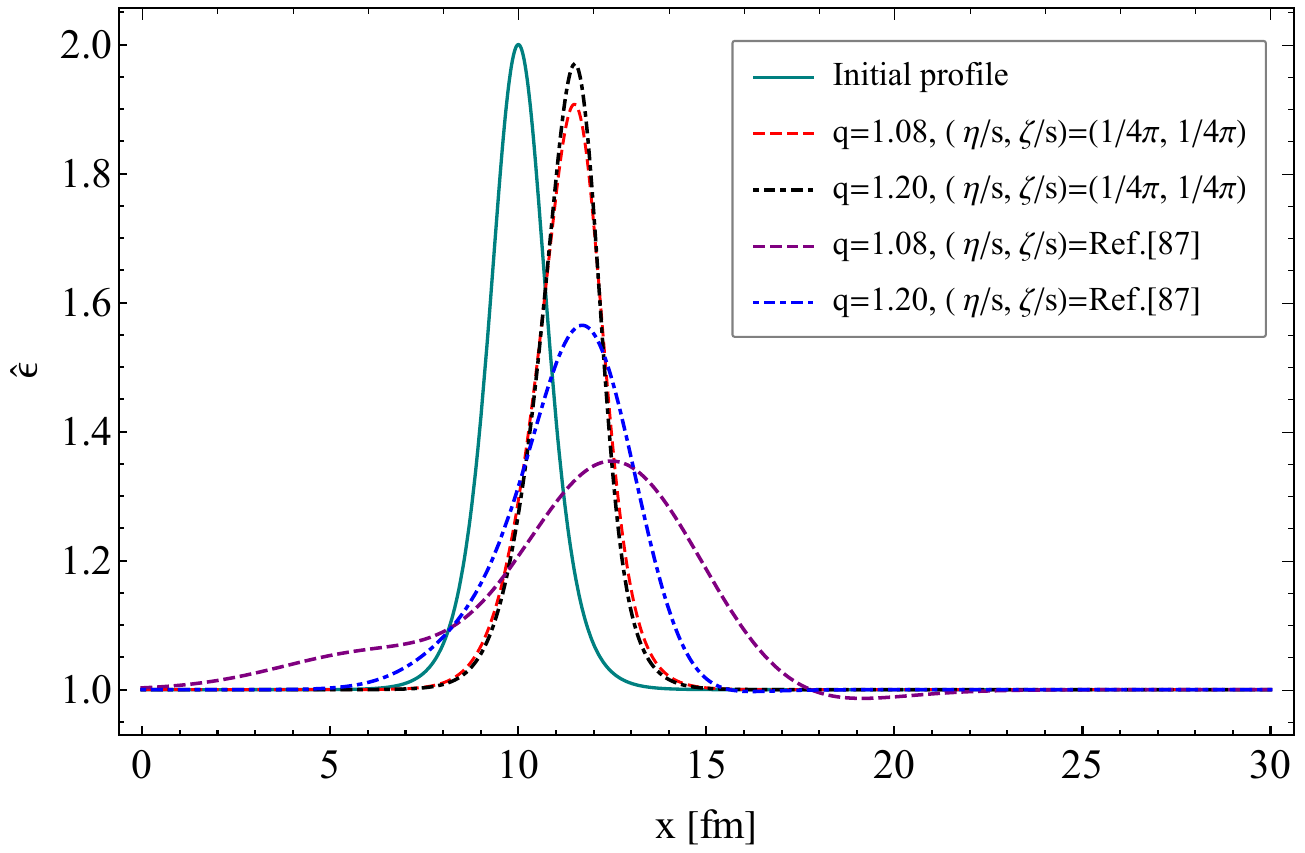}	
\caption{(Color online) The variation of $\hat{\epsilon}$ with spatial coordinate ($x$)
at $t=2$ fm for different values of $q$ and viscous coefficients have been shown here.
The temperature of the fluid,  $T=200$ MeV. The values of  $A_{1}, A_{2}$, $B_{1}$, and $B_{2}$ 
are same as Fig.\ref{fig1}(c).
}
\label{fig7}
\end{figure}

Fig.~\ref{fig5} shows the fate of the same initial perturbation as that of displayed in Fig.~\ref{fig1}(c) after a time of 
$10$ fm for three different values of
$q$ for  $T=200$ MeV. In the presence of dissipation the  $q$ plays an interesting role.
Namely for higher values of $q$  the dissipation is lower and the stiffness is more. In the absence of dissipation (ideal background) we also find
more stiffness for higher $q$~\cite{Bhattacharyya:2020sua}. Indicating that the effects of dissipation can be reduced by increasing $q$, or in other words 
systems with higher $q$ imitate lower dissipation. 
It is also to be noted from the results shown in Fig.~\ref{fig5} that 
the propagation speed of the peak increases with $q$ if 
the $q$ dependence is considered only through EoS. 
The results of ideal $q$-hydrodynamics can be obtained from our results by setting 
the values of the transport coefficients to zero. 
In Fig.~\ref{fig6} the results for vanishingly small transport coefficients have been shown.
In such cases we find that the waves loose their
localization (breaking waves) and the shape preserving nature no longer survives.  

So far we have discussed the effects of non-extensivity on the propagation of nonlinear waves only through the EoS. 
Since the EoS and transport coefficients control the bulk evolution of a system, therefore, for a thorough treatment 
of the propagation of nonlinear waves in a non-extensive background ($q$-background), one should consider the effects of 
non-extensivity, not only through the EoS ($q$-EoS) but also on the transport coefficients ($q$-viscosities) 
which are estimated for $q$-background (see~\cite{Biro:2011bq, Kadam:2015xsa} for details). The results with the $q$-viscosities 
($q$-dependence of viscous coefficients) along with 
the $q$-EoS (as discussed in section~\ref{sec3}) are presented in Fig.~\ref{fig7} (purple and blue dashed lines)
for the initial profile shown by green line.
The results obtained with the effects of $q$-EoS only is compared (black and red dashed lines). It is found that the dissipation 
of nonlinear waves is more when non-extensivity is introduced via the $q$-EoS and the $q$-viscosities together. 
Interestingly, here, the propagation speed of the peak of the perturbations decrease with the increasing value of $q$
when the effects of $q$-viscosity is introduced. 
Along with the reduction of the dissipation the higher values of $q$ also reduces the speed of the propagation. 
This means that the introduction of  
non-extensivity favours the non-dissipative (or anti-dissipative) slowing down of nonlinear perturbations. 
Considering the fact that for an extensive background, the slowing down is caused by the dissipative effects~\cite{Sarwar:2020oux}, 
the results displayed in Fig.~\ref{fig7} reveal that the propagation of non-linear wave  can be useful 
to distinguish between the extensive and the non-extensive background. We also find that 
the dissipation is more for smaller $q$-values (compare with Fig.~\ref{fig5}).  With the q-viscosities along with q-EoS, the relative dissipation with $q$ is more than that of the q-EoS only. This is in contrast to the expectation from the notion that the non-extensive distribution is treated as 
a non-equilibrium deviation from the extensive one. In such a scenario the dissipative correction is interpreted as 
a result of $q\neq 1$. These findings are opposite to the normal expectations. Rather, it indicates that $q$-equilibrium 
should not always be interpreted as a dissipative non-equilibrium form of extensive statistics.

In a system with $q$-background, the higher values of $q$ is accompanied by less dissipation in the system. 
This can be understood from the fact that the enhanced correlation in a system can be 
represented by higher $q$-values~\cite{2020PhRvE.101d0102L,ABE1999403}.  The correlations enhance the fluctuations, 
which can also be represented by a value of $q\neq 1$~\cite{Wilk:1999dr}.  The correlations opposes the entropy 
generation in a system and hence dissipation. Therefore, a higher value of $q$ parameter representing a 
higher degree of correlation which is expected to results in less dissipation, 
this is manifested in the variation of viscosity to entropy ratio ($\eta/s$) with $q$~\cite{Biro:2011bq}. 
The $\eta/s$ decreases with increasing $q$ due to greater rate of change of non-extensive entropy with $q$ 
than of the bulk and shear viscosities, thereby reducing the dissipation.
 
The second and higher order space 
derivatives in the evolution equations contain  the effects of dissipation. 
Without the presence of the viscous terms, the $q$ alone (through $q$-EoS only) can not produce such dissipative nature, 
{\it{i.e}}, $q$ can only have a role in dissipation in the presence of explicit dissipative terms. That is why, without considering 
the $q$-viscosities (or any dissipative effects arising from $d$-hydrodynamics), the authors in~\cite{Bhattacharyya:2020sua}  
do not find any dissipation with $q$-EoS alone. Also they have considered only the first order of energy perturbation 
and found that the equation to be of the breaking-wave type, causing a tilting (ultimately breaking) of nonlinear waves without any dissipation. 
Whereas, in this work we have obtained  KdV type equation by considering perturbations up to second-order with the inclusion of all 
the relevant transport coefficients and found to produce prominent dissipative effects without much tilting.

In such a scenario, one may be tempted to think that, perhaps, less dissipation is the unique signature of presence of 
finite $q$ (increasing $q$-values). However, it may be noted that the less dissipation may be accounted through the 
smaller values of the transport coefficients also. So there may be an ambiguity behind the observed smaller dissipation 
in a system, whether it is a non-extensive system with larger values of $q$ or it is an extensive system with lower viscosity. 
Such  ambiguity can be removed through  other effects which  can 
distinguish these two separate situations. The role of higher values of $q$ as 
the cause of less dissipation can be mimicked by reducing the values of the transport coefficients, but it  may possess 
other distinct effects which will help in removing the ambiguity. In the case of nonlinear wave propagation, the nature of 
the nonlinear wave will provide the relevant distinct feature. With less dissipation in extensive background ({\it{i.e} } without $q$-background)  
the nonlinear wave will not  preserve the solitonic nature due to breaking of the wave (see Fig.~\ref{fig6}), 
whereas the $q$ makes the dissipation less with preserving the solitonic nature (see Fig.~\ref{fig7}). Therefore, the causes for 
less dissipation with Tsallis background can be distinguished with the propagation of a nonlinear wave. This may pave the 
way for investigation to ascertain whether the viscosity of the QGP is really low or it has $q$-thermalized local equilibrium with not 
so low transport coefficients. Since $p+p$ collision admits $q$-distributed particle spectra, therefore, one would argue for 
consideration of $q$-thermalized local equilibrium. In that case it is very important to remove the ambiguity about viscosities 
of QGP estimated by fitting the experimental results. 
Another important point to be noted is that,  when the ideal $q$-hydrodynamics is interpreted as equivalent 
to $d$-hydrodynamics, one should not take $q$ as a cause of dissipation in general sense. This is so because, $q$ does not 
necessarily represents dissipation, especially when the background is in $q$ equilibrium.

\section{Summary and conclusion}
\label{sec5}
In summary, we have derived the equations for the  propagation of nonlinear waves in a dissipative
fluid involving shear viscosity,  bulk viscosity and thermal conductivity. The effect of the $q-$statistics is incorporated by taking into 
account the $q-$EoS and $q$-viscosities. We find that the equation for the second order perturbation is similar to the KdV equation resulting 
in the solitonic nature of nonlinear waves. It is observed that the dissipation of the  wave
is more in first order than the second order.  For larger amplitude of the perturbations, we find that the speed of the 
propagation is larger, whereas for broader perturbations, the dissipation is comparatively less and propagation is faster. 
As far as the background temperature of the medium is concerned, dissipation is less for higher temperature. 
The effect of $q$ is evident from the results shown in Fig.\ref{fig5}, where we see more dissipation of nonlinear waves 
for smaller values of $q$. This is in line with the interpretation of $q$ as representative of correlations as well as 
fluctuations in a system.  The larger values of $q$  may imitate the situation with  smaller values of 
the transport coefficients. 
This is reflected through the  
maintenance of solitonic nature of the nonlinear waves. Moreover, it is found that the non-extensivity favours the non-dissipative slowing down 
of nonlinear perturbations. This is in contrast to the extensive background where the slowing down comes with the 
dissipation~\cite{Sarwar:2020oux}. This may be a useful criterion for  making distinction between the extensive and non-extensive 
background. So the nature of propagation of a nonlinear wave can be potentially used to distinguish a non-extensive medium 
from an extensive one, which could be especially helpful in situations where momentum distribution of particles can not distinguish 
the nature of underlying statistics (Gibbs-Boltzmann or Tsallis). This indistinguishability may occur when there is an uncertainty 
on whether the equilibration 
is complete or not and when there are other sources of power law distributions such as the radiation 
from particles with high momentum
or jets moving through the QGP, instead of the presence of correlations that causes non-extensivity of the medium.

\section{Acknowledgement}
\label{sec8}
 MH, MR would like to thank Department of Higher Education, Govt. of West Bengal, India. TB acknowledges partial support from the joint project between the JINR and IFIN-HH. The work of AB is  supported by Alexander von Humboldt (AvH) foundation and Federal Ministry of Education and Research (Germany)  through Research Group Linkage programme.
\appendix
\section{}
\label{appendixA}
In this appendix we provide the expressions for the relaxation and coupling coefficients~\cite{Israel:1979wp}
required to solve the Israel-Stewart hydrodynamical equations. 
\beqa
\alpha_{0}&=&(D_{41}D_{20}-D_{31}D_{30})\Lambda \phi \Omega J_{21}J_{31}~,\nn\\
\alpha_{1}&=& (J_{41}J_{42}-J_{31}J_{52}) \Lambda \phi J_{21}J_{31}~, \nn\\
\beta_{0}&=& \frac{3\beta}{\phi^{2}\Omega^{2}}[5J_{52}-\frac{3}{D_{20}}\{J_{31}(J_{31}J_{30}-J_{41}J_{20})+J_{41}(J_{41}J_{10}-J_{31}J_{20})\}]~, \nn\\
\beta_{1}&=& \frac{D_{41}}{\Lambda^{2}nmJ_{21}J_{31}}~, \nn\\
\beta_{2}&=& \frac{\beta J_{52}}{2\phi^{2}}~,
\eeqa
where, 
\beqa
D_{rs}&=&J_{r+1,s}J_{r-1,s}-(J_{rs})^{2},\,\,\,\, \phi=T(\epsilon+P),\,\,\,\, 
\psi=\frac{\epsilon +p}{n_im_i}, \nn\\
\Lambda&=& 1+5(\frac{\phi}{nm})-\psi^{2},\,\,\,\, 
\Omega= 3\Big(\frac{\pd ln \phi}{\pd ln n}\Big)-5~,
\eeqa
where $r$, $s$ are integers, $m_i$ and $n_i$ are respectively the mass and density of $i$-th type particle. 
The quantities, $\phi, \psi$ and  $\Lambda$ are calculated by using their relations with  $\epsilon$, $p$ and $n_i$. 
We take the current quark mass of up and down flavors as 10 MeV. 
$J_{rs}$ is defined as,
\beqa
J_{rs}=\frac{A_{0}}{(2s+1)!!}\int_{0}^{\infty} N \Delta Sinh^{2(s+1)}\mathcal{R} Cosh^{r-2s}\mathcal{R} d\mathcal{R}~,
\eeqa
and
\beqa
N=\frac{1}{exp_{q}(\beta Cosh \mathcal{R} -\alpha)-\epsilon}~,
\eeqa
where, $q$-deformed exponential is defined as,
\beqa
exp_{q}(x)=\big[1+(1-q)x \big]^{1/(1-q)}~,
\eeqa
and 
\beqa
\Delta=1+\epsilon N,\,\,\,\, A_{0}=4\pi m_i^3~.
\label{eq14}
\eeqa
These coupling and relaxation coefficients in Landau frame are connected to the coupling and relaxation 
coefficients in the Eckart frame by the following relation~\cite{Israel:1979wp}
\begin{equation}
\tilde{\alpha_{0}}-\alpha_{0}=\tilde{\alpha_{1}}-\alpha_{1}=-(\tilde{\beta_{1}}-\beta_{0})
=-[(\epsilon+p)]^{-1}
\end{equation}

\section{}
\label{appendixB}
The coefficients $\mathcal{S}_i$'s for $i=1$ to $5$ appearing in Eq.~\eqref{nlw2} are given
in this appendix. 
\beqa
\mathcal{S}_{1}&=&\frac{1}{\epsilon_{0}+p_{0}}  \Big[c_s \{\epsilon _0 \left(1- \hat{\epsilon }_1\left(c_s^2-1\right)\right)+p_0\}\Big];\nn\\
 \mathcal{S}_{2}&=& \frac{1}{\epsilon_{0}+p_{0}}  \Big[\epsilon _0 c_s \{c_s^2-1\} \{\epsilon _0 \left(\left(2 c_s^2+1\right) \hat{\epsilon }_1{}^2-\hat{\epsilon }_2\right)-p_0 \hat{\epsilon }_2\}\Big];\nn\\
\mathcal{S}_{3}&=& \frac{1}{12(\epsilon_{0}+p_{0})^{2}}  \Big[\epsilon _0 \hat{\epsilon }_1 \{3 c_s^2 (7 \zeta +8 \eta )+3 \zeta +4 \eta \}\Big];\nn\\
 \mathcal{S}_{4}&=&- \frac{1}{72 c_s c_V (\epsilon _0+p_{0})^2}
\Big[4 c_{V} c_s^2 \{3 \kappa  T \epsilon _0(3 \alpha _0 \zeta +4 \alpha _1 \eta )+3\kappa T(3 \zeta +4 \eta) +3 \kappa T p_0 (3 \alpha _0 \zeta +4 \alpha _1 \eta )\nn\\
&&+(\epsilon _0+p_{0}) (9 \beta _0 \zeta ^2+16 \beta _2 \eta ^2)\}-c_{V}(3 \zeta +4 \eta )^2+12 \kappa  \{\epsilon _0+p_{0}\} \{\epsilon _0 (3 \alpha _0 \zeta +4 \alpha _1 \eta )+3 \zeta +4 \eta \nn\\
&&+p_0 (3 \alpha _0 \zeta +4 \alpha _1 \eta )\}\Big];\nn\\
{\text and}\\
  \mathcal{S}_{5} & =&-\frac{3 \zeta +4 \eta }{6 \left(p_0+\epsilon _0\right)}~.
\eeqa

\section{}
\label{appendixC}
In this appendix the expressions for  $\epsilon_{i,\ell}$ and $P_{i,\ell}$ ($i=F, B$; $\ell=1,2$) are provided as follows.
($i=F, B$; $\ell=1,2$) are given by  \cite{Bhattacharyya:2020sua},
\bea
\epsilon_{B,1} &=& \frac{g}{2 \pi ^2 (q-1)^3 q} \left[ 3 \psi
   ^{(0)}\left(\frac{3}{q}-2\right) + \psi
   ^{(0)}\left(\frac{1}{q}\right) \right.
    \left. - 3 \psi ^{(0)}\left(\frac{2}{q}-1\right) - \psi
   ^{(0)}\left(\frac{4}{q}-3\right) \right], \label{epsilonb1} \\
\epsilon_{F,1} &=& \frac{g}{2 \pi ^2 (q-1)^3 q} \left[3 \Phi \left(-1,1,\frac{2}{q}-1\right)  
-3 \Phi \left(-1,1,\frac{3}{q}-2\right)
+\Phi \left(-1,1,\frac{4}{q}-3\right)    \right. \nn\\ 
&& \left. -\Phi \left(-1,1,\frac{1}{q}\right)\right] \label{epsilonf1}, \\
\epsilon_{F,2} &=& -\frac{g}{4 \pi ^2 (q-1) q} \left[ \Phi \left(-1,1,\frac{2}{q}-1\right) \right. \left.-\Phi
   \left(-1,1,\frac{1}{q}\right)\right], \label{epsilonf2} \\
P_{B,1} &=& \frac{\epsilon_{B,1}}{3},~P_{F,1} =  \frac{\epsilon_{F,1}}{3},~ P_{F,2} = \epsilon_{F,2} \label{pb1pf1pf2}~, 
\eea
where $\psi^{(0)}$ is the digamma function, and $\Phi$ is another special function known as Lerch's transcendent \cite{bateman}.

\bibliography{NonLin_NonExt} 

\end{document}